\documentclass[11pt]{article}
\usepackage[english]{babel}

\begin{document}
\title{Gaussian Random Matrix Ensembles in Phase Space}
\author{Maciej M. Duras $^*$}
\date{$^*$ Institute of Physics, Cracow University of Technology, \\
ulica Podchorazych 1, PL-30-084 Cracow, Poland \\
Electronic address: pfduras @ cyf-kr.edu.pl \\
30th June, 2019}

\maketitle

\begin{abstract}
A new class of Random Matrix Ensembles is introduced.
The Gaussian orthogonal, unitary, and symplectic ensembles
GOE, GUE, and GSE, of random matrices
are analogous to the classical Gibbs ensemble governed by
Boltzmann's distribution in the coordinate space. The proposed new class 
of Random Matrix ensembles is an extension of the above Gaussian ensembles
and it is analogous to the canonical Gibbs ensemble governed
by Maxwell-Boltzmann's distribution in phase space.
The thermodynamical magnitudes of partition function,
intrinsic energy, free energy of Helmholtz, free energy of Gibbs,
enthalpy, as well as entropy, equation
of state, and heat capacities, are derived 
for the new ensemble.
The examples of nonideal gas with quadratic potential energy
as well as ideal gas of quantum matrices are provided.
The distribution function for the new ensembles
is derived from the maximum entropy principle.  
\end{abstract}

\section{Introduction}
\label{sec-introduction}
The Random Matrix Theory RMT is a well established
branch of mathematics and it studies random matrices
(random matrix variables)
defined over following fields ${\bf F}$:
real {\bf R}, complex {\bf C}, and quaternion {\bf H}.
The matrix elements are random variables
with assumed distribution functions.
Many random magnitudes derived from random matrices
are also studied: their eigenvalues, eigenvectors,
eigenphases, determinants, kernels, correlation functions etc.
The application of RMT to nuclear physics began with
works of von Neumann and Wigner 
\cite{von Neumann Wigner 1929,Wigner 1957,Wigner 1956,Porter 1965 Wigners 1 1,Porter 1965 Wigners 1 2} 
and also Landau and Smorodinsky \cite{Landau 1955}. 
They assumed a statistical hypothesis 
for the many-body quantum Hamiltonian 
to explain observed nuclear spectra.
They postulated that the Hamiltonian operator acting in
truncated $N$-dimensional Hilbert space $V_{{\cal X}}={\bf F}^{N}$
is a random matrix with matrix elements Gaussian distributed.
So the definition of $N$-dimensional
Gaussian orthogonal, unitary, and symplectic ensembles
GOE($N$), GUE($N$), GSE($N$), was made,
as well as Poisson ensemble PE.
RMT is different from both classical statistical mechanics
and quantum statistical mechanics. 
In the classical statistical mechanics the approach is based
on dynamics of the considered system in the phase-space.
In the quantum statistical mechanics the studied space is
Fock's space on which the quantum operators act.
Both statistical mechanics deal with three ensembles:
microcanonical, canonical, and grand canonical.
The hermitean random matrices in RMT are derived from the symmetry principle
and not from dynamics. Hence, there are the three classes of ensembles:
orthogonal, unitary, and symplectic, corresponding to three
Lie's groups: orthogonal O($N$, {\bf F}), unitary U($N$, {\bf F}), and symplectic Sp($N$, {\bf F}), that leave invariant
the matrix Haar's measure. However if the random matrices are not hermitean,
then the Lie's group is general linear over field ${\bf F}$, 
thence there are three Ginibre's ensembles. 
The principal difference between statistical mechanics and RMT
is that in the former case the ensemble
consists of systems with the same Hamiltonian
and with different initial condition
whereas in the latter case the matrix ensemble consists of
systems with different quantum matrices, {\it e.g.},
with different quantum Hamiltonians, but with the same symmetry class.
Since the 1950s the march of RMT
through physics was imposing with applications to
nuclear physics, atomic physics, condensed phase physics,
field theory, quantum gravity, quantum chromodynamics,
quantum chaos, disordered mesoscopic systems
\cite{Bohigas 1983,Haq Pandey Bohigas 1982,Bohigas Haq Pandey 1985,Porter Rosenzweig 1960 Suomalaisen,Dyson 1962 140,Dyson 1962 157,Dyson 1962 166,Dyson 1970,Dyson Mehta 1963,Mehta Dyson 1963,Mehta 1971,Pandey 1983,Mehta 1984,Harney 1982,Balian 1968,Gade 1993,Andreev 1994,Haake 1990,Guhr 1998,Mehta 1990 0,Reichl 1992,Bohigas 1991,Porter 1965,Brody 1981,Beenakker 1997}. 

The aim of the present article 
is an extension of the Gaussian ensembles of RMT
to a wider class of ensembles in matrix phase space.
The ensembles (Gaussian or non-Gaussian) of RMT
deal with quantum random matrices from 
the quantum matrix space analogous to the configuration space of
classical statistical mechanics. The proposed extension will lead one
the ensembles of pairs of random quantum matrices from
the quantum matrix space analogous to the phase-space of classical
statistical mechanics. 
As far as we consider the Gaussian ensembles of RMT,
the probability density function $f_{{\cal X}}$ of random quantum Hamiltonian
matrix $X$ belonging to GOE($N$), GUE($N$), or GSE($N$) reads \cite{Guhr 1998}:
\begin{eqnarray}
& & f_{{\cal X}}(X)={\cal C}_{X \beta} 
\exp{[-\beta \cdot \frac{1}{2} \cdot {\rm Tr} (X^{\dag}X)]},
\label{pdf-GOE-GUE-GSE} \\
& & {\cal C}_{X \beta}=(\frac{\beta}{2 \pi})^{{\cal N}_{X \beta}/2}, 
\nonumber \\
& & {\cal N}_{X \beta}=N+\frac{1}{2}N(N-1)D, \nonumber
\end{eqnarray}
where $\beta$ is an inverse of the temperature $T$ measured in
energetic scale, $D$ is dimension of random matrix elements, 
the parameters $\beta, D$, assume values 
$\beta=1,2,4,$ and $D=1,2,4,$ for GOE($N$), GUE($N$), GSE($N$), respectively,
and ${\cal N}_{X \beta}$ is number of independent matrix elements
of hermitean Hamiltonian $X=X^{\dag}$.
The quantum Hamiltonian $X$ is a random matrix variable and
it is zero-centred Gaussian distributed with the diagonal
covariance matrix 
${\rm Cov} (X_{ij}, X_{kl})= \frac{1}{\beta} \delta_{ik} \delta_{jl}$.
The matrix elements $X_{ij}$ are independently Gaussian distributed,
and $X_{ij} \in {\bf R}, {\bf C}, {\bf H}$, 
for GOE, GUE, GSE, respectively.
The normalization of distribution of $X$ is:
\begin{eqnarray}
& & \int f_{{\cal X}}(X) dX=1,
\label{pdf-GOE-GUE-GSE-normalization} \\
& & dX=\prod_{i=1}^{N} \prod_{j \geq i}^{N} 
\prod_{\gamma=0}^{D-1} dX_{ij}^{(\gamma)}, \nonumber \\
& & X_{ij}=(X_{ij}^{(0)}, ..., X_{ij}^{(D-1)}) \in {\bf F}, \nonumber
\end{eqnarray}
where $dX$ is Haar's measure in the matrix space.
The Haar's measure $dX$ is invariant under transformations
from the orthogonal O($N$, {\bf F}), unitary U($N$, {\bf F}), 
and symplectic Sp($N$, {\bf F}) Lie's groups of symmetries, respectively. 
The probability distribution $f_{{\cal X}}$
is invariant under the three Lie's groups, respectively.
The Hamiltonian operators $X$ act in given Hilbert space
of eigenvectors 
$V_{{\cal X}}={\bf F}^{N}$,
so they belong themselves to the Hilbert space
$W_{{\cal X}}={\rm Herm}({\bf F}, N)$
of all $N \times N$ hermitean matrices
with matrix elements belonging to the field $\bf F$. 
In the space $W_{{\cal X}}$
the scalar product of two operators $X_{1}, X_{2}$,
is given by formula 
\begin{equation}
\left< X_{1} | X_{2} \right> = {\rm Tr} (X_{1}^{\dag} X_{2}),
\label{scalar-product-X}
\end{equation}
which yields
\begin{equation}
\left< X_{1} | X_{2} \right> = {\rm Tr} (X_{1} X_{2}),
\label{scalar-product-X-hermitean}
\end{equation}
because $X_{1}^{\dag}=X_{1}$.
Since the space $W_{{\cal X}}$ 
is also Banach space, then the norm of operator $X$ reads:
\begin{equation}
|| X || = ( | \left< X | X \right> | )^{1/2}
= ( | {\rm Tr} (X^{\dag} X) | )^{1/2}.
\label{norm-X}
\end{equation}
Due to the hermiticity of $X$
we have property $\left< X | X \right> \geq 0$,
and then 
\begin{equation}
|| X || = (\left< X | X \right>)^{1/2}= [ {\rm Tr} (X^{2}) ]^{1/2}.
\label{norm-X-hermitean}
\end{equation}
It follows that $|| X ||$ is Euclidean norm, and
the distance of two matrices $X_{1}, X_{2}$,
is given by 
\begin{equation}
d(X_{1}, X_{2})= || X_{1}-X_{2} ||.
\label{distance-X}
\end{equation}
Using Eqs. 
(\ref{scalar-product-X}), (\ref{scalar-product-X-hermitean}),
(\ref{norm-X}), (\ref{norm-X-hermitean}), 
we rewrite the distribution Eq. (\ref{pdf-GOE-GUE-GSE}): 
\begin{equation}
f_{{\cal X}}(X)={\cal C}_{X \beta} 
\exp{[-\beta \cdot \frac{1}{2} \cdot || X ||^{2}]}.
\label{pdf-GOE-GUE-GSE-norm} 
\end{equation}
From the above properties of scalar product, norm, and distance,
we infer the analogy between configuration space ${\bf F}^{N}$ of
$x$-components of generalized coordinates of any one-dimensional 
classical system of $N$ particles,
and the configuration space $W_{{\cal X}}$ of generalized ''$X$-coordinate''
of quantum system described by Hamiltonian operator $X$.
From the formulae 
(\ref{pdf-GOE-GUE-GSE-normalization}),(\ref{pdf-GOE-GUE-GSE-norm}),
we deduce that the quantum Hamiltonian $X$ 
has continuous (non-discrete) distribution
that is analogous to the distribution of x-coordinate 
of one-dimensional classical particle in
potential of harmonic oscillator $U(x)=\frac{1}{2}x^{2}$.
The $x$-coordinate of the classical particle 
has the Boltzmann's distribution
\cite{Feynman 1972,Tolman 1967,Huang 1963,Anselm 1973,Gyarmati 1970,Zubaryev 1971,Fowler 1936,Fowler 1965,Hill 1956,Landau 1951,Toda 1983,Balescu 1975,Reiff 1965,Balian 1982}:
\begin{equation}
f_{{\cal B}}(x)={\cal C}_{X} 
\exp{[-\frac{1}{k_{B} T} \cdot \frac{1}{2} k x^{2}]},
\label{pdf-Boltzmann} 
\end{equation}
The classical particle's statistics is governed by canonical ensemble
in configuration space (coordinate space).
In view of this analogy parameter $\beta$ corresponds
to $\frac{1}{k_{B} T}$.
Hence, the discrete "temperature" $T$ for the Gaussian ensembles
reads:
\begin{equation}
T=\frac{1}{k_{B} \beta}.
\label{temperature-beta} 
\end{equation}
The main point is that the quantum Hamiltonians $X$
are governed by classical continuous distribution
(\ref{pdf-GOE-GUE-GSE-normalization}),(\ref{pdf-GOE-GUE-GSE-norm}),
and not by quantum discrete distribution.
The quantum Hamiltonians are belonging to the configuration space
(generalized coordinates's space) 
$W_{{\cal X}}$. The Gaussian ensembles GOE, GUE, GSE, 
of the Hamiltonians
describe the classical non-ideal gas of harmonic oscillators
of quantum matrices in configuration space.
We naturally extend this description by the introduction 
of momentum space and phase space for quantum Hamiltonians.
Firstly, we concentrate on definition of momentum space of random matrices
corresponding to configuration space.
We define the quantum operator of generalized linear momentum $P$
that is correlated with  
the quantum operator $X$ of generalized ``$X$-positon'' (``$X$-coordinate'').
The classical observables of generalized momentum
and generalized position are classically canonically conjugated
in classical mechanics and in classical statistical mechanics.
The operator $P$ acts on Hilbert space 
$V_{{\cal P}}={\bf F}^{N}$,
and it belongs to Hilbert space $W_{{\cal P}}={\rm Herm}({\bf F}, N)$
of momenta.
The scalar product, norm, and distance in $W_{{\cal P}}$
are as follows:
\begin{eqnarray}
& & \left< P_{1} | P_{2} \right> = {\rm Tr} (P_{1}^{\dag} P_{2}),
\label{scalar-product-P} \\
& & || P || = ( | \left< P | P \right> | )^{1/2}
= ( | {\rm Tr} (P^{\dag} P) | )^{1/2}.
\label{norm-P} \\
& & d(P_{1}, P_{2})= || P_{1}-P_{2} ||.
\label{distance-P}
\end{eqnarray}
We assume that $P$ is hermitean $P^{\dag}=P$,
which yields $\left< P | P \right>  \geq 0$,
and 
\begin{equation}
|| P || = (\left< P | P \right>)^{1/2}= [ {\rm Tr} (P^{2}) ]^{1/2}.
\label{norm-P-hermitean}
\end{equation}
Again, the above properties of scalar product, norm, and distance
in momentum space,
conduct us to the analogy between momentum space ${\bf F}^{N}$ of
$x$-components of generalized linear momenta 
of any one-dimensional classical system of $N$ particles,
and the momentum space $W_{{\cal P}}$ of ''$P$-momenta''
of quantum system described by momentum operator $P$.
The postulated distribution of momentum is analogous
to classical Maxwell's distribution and it reads:
\begin{eqnarray}
& & f_{{\cal P}}(P)={\cal C}_{P \beta} 
\exp{[-\beta \cdot \frac{1}{2 M} \cdot {\rm Tr} (P^{\dag}P)]},
\label{pdf-GOE-GUE-GSE-P} \\
& & {\cal C}_{P \beta}=(\frac{\beta}{2 \pi M})^{{\cal N}_{P \beta}/2}, 
\nonumber \\
& & {\cal N}_{P \beta}=N+\frac{1}{2}N(N-1)D, \nonumber \\
& & \int f_{{\cal P}}(P) dP=1,
\label{pdf-GOE-GUE-GSE-P-normalization} \\
& & dP=\prod_{k=1}^{N} \prod_{l \geq k}^{N} 
\prod_{\delta=0}^{D-1} dP_{kl}^{(\delta)}, \nonumber \\
& & P_{kl}=(P_{kl}^{(0)}, ..., P_{kl}^{(D-1)}) \in {\bf F}, \nonumber
\end{eqnarray}
where $M$ is "mass" of the particle in matrix space.
The Haar's measure $dP$ is invariant under transformations
from the orthogonal O($N$, {\bf F}), unitary U($N$, {\bf F}), 
and symplectic Sp($N$, {\bf F}) Lie's groups of symmetries, respectively.
Also the probability density function $f_{{\cal P}}$ is invariant under
above Lie's groups.
Formally momentum $P$ belongs to Gaussian ensembles that we
will denote by GOE($N$, $P$), GUE($N$, $P$), GSE($N$, $P$).
Momentum is zero-centred Gaussian distributed
with diagonal covariance matrix 
${\rm Cov} (P_{kl}, P_{mn})= \frac{M}{\beta} \delta_{km} \delta_{ln}$. 
Secondly, we are able now to introduce a phase space $W_{\Gamma}$ of
generalized canonically conjugated operators 
of linear momenta $P$ and ``$X$-coordinates''.
The phase space $W_{\Gamma}$ is an {\em analogy} to
classical phase space ${\bf F}^{2N}$ of $x$-components
of generalized coordinates 
and generalized linear momenta of classical one-dimensional
system of $N$ particles.
The pair of operators $(X, P)=\Gamma$ (the direct sum of operators)
composes a point
in phase space of random matrices 
$W_{\Gamma} = W_{{\cal X}} \times W_{{\cal P}}$.
The Haar's measure $d \Gamma$ in the matrix phase space is given by:
\begin{eqnarray}
& & d \Gamma = 
\frac{1}{{\cal N}_{\Gamma \beta}! 
\cdot h^{s_{\Gamma \beta}}} \cdot dX \cdot dP,
\label{measure-HP} \\
& & {\cal N}_{\Gamma \beta}=
{\cal N}_{P \beta} = {\cal N}_{X \beta}
=N+\frac{1}{2}N(N-1)D, \nonumber \\
& & s_{\Gamma \beta}={\cal N}_{\Gamma \beta}. \nonumber
\end{eqnarray}
The Haar's measure $d \Gamma$ is invariant 
under composite transformations $U_{\Gamma}=(U_{X},U_{P})$
from the direct sums of the Lie's groups of symmetries:
orthogonal ${\rm O}(N, {\bf F}) \oplus {\rm O}(N, {\bf F})$, 
unitary ${\rm U}(N, {\bf F}) \oplus {\rm U}(N, {\bf F})$, 
and symplectic ${\rm Sp}(N, {\bf F}) \oplus {\rm Sp}(N, {\bf F})$, respectively.
The distribution of the pair $\Gamma$ is postulated in
the following form:
\begin{eqnarray}
& & f_{\Gamma}(\Gamma)={\cal C}_{P \beta} 
\exp{[-\beta \cdot (\frac{1}{2 M} \cdot {\rm Tr} (P^{\dag}P)
+ \frac{K}{2} \cdot {\rm Tr} (X^{\dag}X) )]}=
\label{pdf-GOE-GUE-GSE-HP} \\
& & ={\cal C}_{P \beta} 
\exp{[-\beta \cdot (\frac{1}{2 M} \cdot || P ||^{2}
+ \frac{K}{2} \cdot || X ||^{2} )]}, \nonumber \\
& & {\cal C}_{\Gamma \beta}=
{\cal C}_{P \beta} \cdot {\cal C}_{X \beta}
=(\frac{\beta}{2 \pi M})^{{\cal N}_{\Gamma \beta}} 
\cdot (\frac{K}{M})^{{\cal N}_{\Gamma \beta}/2}, \nonumber \\
& & \int f_{\Gamma}(X, P) dX \cdot dP=1, \nonumber
\end{eqnarray}
which is analog of classical Maxwell-Boltzmann's distribution.
The above distribution is also invariant under direct sums
of orthogonal, unitary, and symplectic Lie's groups of transformations
of composite symmetry.
Hence, we extended both the random quantum matrices $X$ to
direct sums $\Gamma$ of random quantum matrices 
and the symmetry Lie's groups to the direct sums of symmetry Lie's groups.
The Hamiltonian operators $X$ and momentum operators $P$
are independent random variables.
We denote the Gaussian orthogonal, unitary, and symplectic
ensembles in the phase space as follows: 
GOE($N$, $\Gamma$), GUE($N$, $\Gamma$), GSE($N$, $\Gamma$),
whereas the standard Gaussian ensembles in configuration space
might be symbolized by 
GOE($N$, $X$)=GOE($N$), 
GUE($N$, $X$)=GUE($N$), 
GSE($N$, $X$)=GSE($N$), respectively. 

\section{The Thermodynamics of New Ensembles}
\label{sec-thermodynamics}
The "classical" Hamiltonian ${\cal H}_{\Gamma}(X, P)$,
in the matrix phase space is a sum of the "classical" kinetic energy
${\cal T}_{\Gamma}(P)$, 
and the "classical" potential energy ${\cal U}_{\Gamma}(X, P)$:
\begin{equation}
{\cal H}_{\Gamma}(X, P)=
{\cal T}_{\Gamma}(P)+{\cal U}_{\Gamma}(X, P).
\label{classical-H}
\end{equation}
Firstly, let us consider the example of nonideal gas
of harmonic oscillators in the matrix phase space.
Then, $f_{\Gamma}(\Gamma)$ is Maxwell-Boltzmann's
distribution with quadratic potential energy,
and the considered three ensembles of pairs of random matrices are
GOE($N$, $\Gamma$), GUE($N$, $\Gamma$), GSE($N$, $\Gamma$).
The distribution $f_{\Gamma}(\Gamma)$ 
Eq.(\ref{pdf-GOE-GUE-GSE-HP}),
and the ensemble average $\left< g(\Gamma) \right>$ 
of magnitude $g(\Gamma)$ can be finally written in traditional form:
\begin{eqnarray}
& & f_{\Gamma}(\Gamma)={\cal C}_{\Gamma \beta}
\cdot \exp{[-\beta \cdot {\cal H}_{\Gamma}(\Gamma)]},
\label{pdf-GOE-GUE-GSE-HP-partition-function} \\
& & \int f_{\Gamma}(\Gamma) dX \cdot dP =1, 
\nonumber \\
& & \left< g(\Gamma) \right>=
\int g(\Gamma) 
\cdot \exp{[-\beta \cdot {\cal H}_{\Gamma}(\Gamma)]}d \Gamma
/
\int \exp{[-\beta \cdot {\cal H}_{\Gamma}(\Gamma)]}d \Gamma \nonumber \\
& & =
\frac{1}{Z_{\beta}} \cdot
\int \exp{[-\beta \cdot {\cal H}_{\Gamma}(\Gamma)]}d \Gamma, \nonumber
\end{eqnarray}
under following conditions:
\begin{eqnarray}
& & {\cal T}_{\Gamma}(P)=\frac{1}{2 M} \cdot || P ||^{2},
\label{kinetic-potential-energy-HP} \\
& & {\cal U}_{\Gamma}(\Gamma)=\frac{K}{2} \cdot || X ||^{2},
\nonumber \\
& & Z_{\beta}= 
\int \exp{[-\beta \cdot {\cal H}_{\Gamma}(\Gamma)]} d \Gamma=
({\cal N}_{\Gamma \beta}! \cdot h^{s_{\Gamma \beta}} 
\cdot {\cal C}_{\Gamma \beta})^{-1}.
\nonumber
\end{eqnarray}
The partition function $Z_{\beta}$ for the new ensembles
GOE($N$, $\Gamma$), GUE($N$, $\Gamma$), GSE($N$, $\Gamma$),
can be easily calculated and it reads:
\begin{equation}
Z_{\beta}=({\cal N}_{\Gamma \beta}! \cdot h^{s_{\Gamma \beta}})^{-1}
\cdot (\frac{2 \pi}{\beta})^{{\cal N}_{\Gamma \beta}} 
\cdot (\frac{K}{M})^{{\cal N}_{\Gamma \beta}/2}.
\label{partition-function-HP}
\end{equation}
It follows that the Helmholtz's free energy $F_{\beta}$ is equal to:
\begin{equation}
F_{\beta}=-\frac{1}{\beta}
\ln Z_{\beta}
=-\frac{1}{\beta}
\ln [
({\cal N}_{\Gamma \beta}! \cdot h^{s_{\Gamma \beta}})^{-1}
\cdot (\frac{2 \pi}{\beta})^{{\cal N}_{\Gamma \beta}} 
\cdot (\frac{K}{M})^{{\cal N}_{\Gamma \beta}/2}].
\label{free-energy-HP}
\end{equation}
The free energy does not depend on the volume ${\cal V}_{X}$, 
hence the pressure vanishes:
\begin{equation}
p_{\beta}=-\frac{\partial F_{\beta}}{\partial {\cal V}_{X}}=0,
\label{pressure-HP}
\end{equation}
and Eq. (\ref{pressure-HP}) is equation of state,
where ${\cal V}_{X}= \int dX$ is the volume of the subset of
$X$-coordinate space to which the Hamiltonians $X$ are confined.
Entropy of the system $S_{\beta}$ is is proportional 
to Boltzmann's ${\rm H}$ function of the condition of the system 
and it is the ensemble average of
negative $\eta_{\beta}$ of phase operator:
\begin{eqnarray}
& & S_{\beta}= -k_{B} {\rm H} =
\left< \eta_{\beta} \right>,
\label{entropy-eta-HP} \\
& & \eta_{\beta}= 
-k_{B} \ln f_{\Gamma}(\Gamma),
\nonumber \\
& & S_{\beta}=
\int (- k_{B} \ln f_{\Gamma}(\Gamma)) f_{\Gamma}(\Gamma) d \Gamma
= k_{B} \ln Z_{\beta}^{{\rm ID}}+
k_{B} {\cal N}_{\Gamma \beta}.
\nonumber 
\end{eqnarray}
It follows that the intrinsic energy $U_{\beta}$ is equal to:
\begin{equation}
U_{\beta}=\left< {\cal H}_{\Gamma} \right>
=F_{\beta} + T  S_{\beta}=
F_{\beta} + \frac{1}{\beta k_{B}} S_{\beta}=
\frac{{\cal N}_{\Gamma \beta}}{\beta}.
\label{intrinsic-energy-HP} 
\end{equation}
The enthalpy $H_{X\beta}=U_{\beta}+p_{\beta}V$,
and Gibbs's free energy $G_{\beta}=F_{\beta}+p_{\beta}V$,
are given by formulae:
\begin{equation}
H_{X\beta}=U_{\beta},
G_{\beta}=F_{\beta}.
\label{enthalpy-Gibbs-free-HP} 
\end{equation}
The averaged square $\left< {\cal H}_{\Gamma}^{2} \right>$ 
of "classical" energy 
and the variance ${\rm Var} ({\cal H}_{\Gamma})$ read:
\begin{eqnarray}
& & \left< {\cal H}_{\Gamma}^{2} \right>=
{\cal N}_{\Gamma \beta}
\cdot ( {\cal N}_{\Gamma \beta} + 1)
\cdot \beta^{-2},
\label{square-energy-average-HP} \nonumber \\
& & {\rm Var} ({\cal H}_{\Gamma})=
\left< {\cal H}_{\Gamma}^{2} \right> -
\left< {\cal H}_{\Gamma} \right>^{2}=
{\cal N}_{\Gamma \beta} \cdot \beta^{-2}.
\nonumber 
\end{eqnarray}
Finally, 
the heat capacity at constant pressure $C_{p}$,
the heat capacity at constant volume $C_{V}$, 
and isentropic exponent (polytropic exponent) $\kappa$ are:
\begin{eqnarray}
& & C_{p}={\cal N}_{\Gamma \beta} \cdot k_{B}
\label{heat-capacity-p-HP}, \\
& & C_{V}= {\cal N}_{\Gamma \beta} \cdot k_{B},
\label{heat-capacity-V-HP} \nonumber \\
& & \kappa= \frac{C_{p}}{C_{V}}=1.
\nonumber 
\end{eqnarray}

Secondly, we study the example of ideal gas in the
phase space $W_{\Gamma}$ of pairs of quantum random matrices.
For that case 
the "classical" potential energy ${\cal U}_{\Gamma}^{{\rm ID}}(X, P)$ vanishes:
\begin{equation}
{\cal H}_{\Gamma}^{{\rm ID}}(X, P)={\cal T}_{\Gamma}^{{\rm ID}}(P), 
{\cal U}_{\Gamma}^{{\rm ID}}(X, P)=0.
\label{classical-H-HP-ideal}
\end{equation}
The three ensembles with the
"classical" Hamiltonian given by Eq. (\ref{classical-H-HP-ideal}) 
will be denoted as follows 
IDEAL($N$, $\beta$, $\Gamma$), IDEAL($N$, $\beta$, $\Gamma$),
IDEAL($N$, $\beta$, $\Gamma$). 
Hence, $f_{\Gamma}(\Gamma)$ 
Eq.(\ref{pdf-GOE-GUE-GSE-HP-partition-function}) is Maxwell-Boltzmann's
distribution with vanishing potential energy:
\begin{eqnarray}
& & f_{\Gamma}^{{\rm ID}}(\Gamma)={\cal C}_{\Gamma \beta}
\cdot \exp{[-\beta \cdot {\cal H}_{\Gamma}^{{\rm ID}}(\Gamma)]}=
{\cal C}_{\Gamma \beta}
\cdot \exp{[-\beta \cdot {\cal T}_{\Gamma}(\Gamma)]}=
\label{pdf-GOE-GUE-GSE-HP-partition-function-ideal} \\
& & = {\cal C}_{\Gamma \beta}
\cdot \exp{[-\beta \cdot \frac{1}{2 M} \cdot || P ||^{2}]} , \nonumber \\
& & \int f_{\Gamma}^{{\rm ID}}(\Gamma) dX \cdot dP =1. \nonumber 
\end{eqnarray}
The partition function $Z_{\beta}^{{\rm ID}}$ 
for the ideal gas reads:
\begin{equation}
Z_{\beta}^{{\rm ID}}=
({\cal N}_{\Gamma \beta}! \cdot h^{s_{\Gamma \beta}})^{-1}
\cdot (\frac{2 \pi M}{\beta})^{{\cal N}_{\Gamma \beta}/2} 
\cdot {\cal V}_{X}^{{\cal N}_{\Gamma \beta}}.
\label{partition-function-HP-ideal}
\end{equation}
It implies that the Helmholtz's free energy $F_{\beta}^{{\rm ID}}$ equals:
\begin{equation}
F_{\beta}^{{\rm ID}}=-\frac{1}{\beta}
\ln Z_{\beta}^{{\rm ID}}
=-\frac{1}{\beta}
\ln [
({\cal N}_{\Gamma \beta}! \cdot h^{s_{\Gamma \beta}})^{-1}
\cdot (\frac{2 \pi M}{\beta})^{{\cal N}_{\Gamma \beta}/2} 
\cdot {\cal V}_{X}^{{\cal N}_{\Gamma \beta}}
].
\label{free-energy-HP-ideal}
\end{equation}
For the ideal gas, the free energy depends on the volume ${\cal V}_{X}$, 
hence the pressure is:
\begin{equation}
p_{\beta}^{{\rm ID}}=
-\frac{\partial F_{\beta}^{{\rm ID}}}{\partial {\cal V}_{X}}=
\frac{1}{\beta {\cal V}_{X}} \cdot {\cal N}_{\Gamma \beta},
\label{pressure-HP-ideal}
\end{equation}
and Eq. (\ref{pressure-HP-ideal}) is equation of state.
We observe that the entropy of the gas $S_{\beta}^{{\rm ID}}$ is:
\begin{equation}
S_{\beta}^{{\rm ID}}
= k_{B} \ln Z_{\beta}^{{\rm ID}}+
k_{B} \frac{1}{2} {\cal N}_{\Gamma \beta}.
\label{entropy-eta-HP-ideal} 
\end{equation}
These results readily lead 
to the formula for intrinsic energy $U_{\beta}^{{\rm ID}}$:
\begin{equation}
U_{\beta}^{{\rm ID}}=
\frac{{\cal N}_{\Gamma \beta}}{2 \beta}.
\label{intrinsic-energy-HP-ideal} 
\end{equation}
Consequently, the enthalpy $H_{X\beta}^{{\rm ID}}$,
and Gibbs's free energy $G_{\beta}^{{\rm ID}}$,
are given by formulae:
\begin{equation}
H_{X\beta}^{{\rm ID}}=\frac{3 {\cal N}_{\Gamma \beta}}{2 \beta},
G_{\beta}^{{\rm ID}}=\frac{1}{\beta}
\cdot [-\ln Z_{\beta}^{{\rm ID}} + {\cal N}_{\Gamma \beta}].
\label{enthalpy-Gibbs-free-HP-ideal} 
\end{equation}
Clearly, the averaged square 
$\left< ({\cal H}_{\Gamma}^{{\rm ID}})^{2} \right>$ of "classical" energy 
and the variance ${\rm Var} ({\cal H}_{\Gamma}^{{\rm ID}})$ are:
\begin{eqnarray}
& & \left< ({\cal H}_{\Gamma}^{{\rm ID}})^{2} \right>=
\frac{1}{2} \cdot {\cal N}_{\Gamma \beta}
\cdot ( \frac{1}{2} \cdot {\cal N}_{\Gamma \beta} + 1)
\cdot \beta^{-2},
\label{square-energy-average-HP-ideal} \nonumber \\
& & {\rm Var} ({\cal H}_{\Gamma}^{{\rm ID}})=
\frac{1}{2} \cdot{\cal N}_{\Gamma \beta} \cdot \beta^{-2}.
\nonumber 
\end{eqnarray}
Immediately, we have that 
the heat capacities $C_{p}^{{\rm ID}}$, $C_{V}^{{\rm ID}}$, 
and isentropic exponent $\kappa^{{\rm ID}}$ are:
\begin{eqnarray}
& & C_{p}^{{\rm ID}}=
\frac{3}{2} {\cal N}_{\Gamma \beta} \cdot k_{B}
\label{heat-capacity-p-HP-ideal}, \\
& & C_{V}^{{\rm ID}}= 
\frac{1}{2} \cdot {\cal N}_{\Gamma \beta} \cdot k_{B},
\label{heat-capacity-V-HP-ideal} \nonumber \\
& & \kappa^{{\rm ID}}= 3.
\nonumber 
\end{eqnarray}

\section{The Maximum Entropy Principle}
\label{sec-maximal-entropy}
In order to derive the probability distribution
in matrix phase space $W_{\Gamma}$ we apply
the maximum entropy principle:
\begin{equation}
{\rm max} \{S_{\beta}(f_{\Gamma}): \left< 1 \right>=1, 
\left< {\cal H}_{\Gamma} \right>=U_{\beta} \},
\label{maximal-entropy-problem}
\end{equation}
which yields:
\begin{equation}
{\rm max} \{\int (- k_{B} \ln f_{\Gamma}(\Gamma)) f_{\Gamma}(\Gamma) d \Gamma: 
\int f_{\Gamma}(\Gamma) d \Gamma=1,
\int {\cal H}_{\Gamma}(\Gamma) f_{\Gamma}(\Gamma) d \Gamma=U_{\beta} \},
\label{maximal-entropy-problem-equivalent}
\end{equation}
The maximization of entropy $S_{\beta}$ under two additional
constraints of normalization of the probability density function,
and of equality of its first momentum and intrinsic energy,
is equivalent to the minimization of the following functional
${\cal F}(f_{\Gamma})$ with the use of Lagrange multipliers 
$\alpha_{1}, \beta_{1}$: 
\begin{eqnarray}
& & {\rm min} \{ {\cal F} (f_{\Gamma}) \}, 
 \label{maximal-entropy-problem-Lagrange} \\
& & {\cal F} (f_{\Gamma}) 
= \int ( k_{B} \ln f_{\Gamma}(\Gamma)) f_{\Gamma}(\Gamma) d \Gamma
+\alpha_{1} \int f_{\Gamma}(\Gamma) d \Gamma
+ \beta_{1} \int {\cal H}_{\Gamma}(\Gamma) f_{\Gamma}(\Gamma) d \Gamma. 
\nonumber
\end{eqnarray}
It follows, that the first variational derivative of ${\cal F}(f_{\Gamma})$
must vanish:
\begin{equation}
\frac{\delta {\cal F} (f_{\Gamma})}{\delta f_{\Gamma}}=0,
\label{Lagrange-first-derivative}
\end{equation}
which produces:
\begin{equation}
k_{B} (\ln f_{\Gamma}(\Gamma) + 1) 
+\alpha_{1} + \beta_{1} {\cal H}_{\Gamma}(\Gamma)=0,
\label{Lagrange-integrand}
\end{equation}
and equivalently:
\begin{eqnarray}
& & f_{\Gamma}(\Gamma)={\cal C}_{\Gamma \beta}  \cdot
\exp{[-\beta \cdot {\cal H}_{\Gamma}(\Gamma)]}
\label{pdf-GOE-GUE-GSE-HP-partition-function-Lagrange} \\
& & {\cal C}_{\Gamma \beta}= \exp[ -(\alpha_{1}+1) \cdot k_{B}^{-1}],
\beta=\beta_{1} \cdot k_{B}^{-1}.
\nonumber
\end{eqnarray}
The variational principle of maximum entropy does not
force additional condition on functional form 
of energy ${\cal H}_{\Gamma}(\Gamma)$.
Therefore, the distribution 
Eq. (\ref{pdf-GOE-GUE-GSE-HP-partition-function-Lagrange})
defines a very large class of random matrix ensembles 
in phase space of generalized matrix coordinates and matrix momenta.
The $\beta$ parameter can assume any value. 
We can perform threefold restriction: either $\beta$ is equal to 1, 2, 4, 
or ${\cal H}_{\Gamma}(\Gamma)$ is given by 
Eqs (\ref{classical-H}), (\ref{kinetic-potential-energy-HP}),
of finally both conditions are fulfilled.
In the latter case we regain new Gaussian ensembles in phase space
Eq. (\ref{pdf-GOE-GUE-GSE-HP}).
In order to conclude, the derivation of the probability density function 
Eq. (\ref{pdf-GOE-GUE-GSE-HP-partition-function-Lagrange})
is a new approach in Random Matrix Theory, since
it defines a huge class of ensembles of direct sums of quantum operators
of generalized coordinates and momenta in the new matrix phase space
which are distributed according to classical continuous probability density.
The ordinary Lie's groups of symmetries of both the probability densities
and of the Haar's measures are extended to the direct sums 
of Lie's groups of symmetries. 
The studied new ensembles of random matrices describe one-dimensional
nonideal gas with quadratic potential of quantum operators 
and ideal gas of quantum operators.

\section{Acknowledgements}
\label{sect-acknowledgements}
It is my pleasure to deepestly thank Professor Antoni Ostoja-Gajewski
for his continuous help.


\begin{thebibliography}{10} 

\bibitem{von Neumann Wigner 1929}
 J. von Neumann and E. P. Wigner, Phys. Z. {\bf 30}, 462 (1929).
\bibitem{Wigner 1957} 
 E. P. Wigner, 
 International Conference on the
 Neutron Interactions with the Nucleus, 
 Columbia University, New York, September 9-13, 1957,
 Columbia University Report No. CU-175 (TID-7547), 1957, p. 49.
\bibitem{Wigner 1956}
 E. P. Wigner, 
 Conference on Neutron Physics by 
 Time-of-Flight, Gatlinburg, Tennessee, 
 November 1 and 2, 1956, Oak Ridge National Laboratory Report No. ORNL-2309, 
 1957, p. 59.
\bibitem{Porter 1965 Wigners 1 1}
 C. E. Porter, 
 {\em Statistical Theories of Spectra: Fluctuations} 
 (Academic Press, New York, 1965), p. 223.
\bibitem{Porter 1965 Wigners 1 2}
 C. E. Porter, 
 {\em Statistical Theories of Spectra: Fluctuations} 
 (Academic Press, New York, 1965), p. 199. 
\bibitem{Landau 1955}
 L. Landau and Ya. Smorodinsky, 
 {\em Lectures on the Theory of the Atomic Nucleus } 
 (State Technical - Theoretical Literature Press, Moscow, 1955), 
 (translation: Consultants Bureau, Inc., New York, 1958, p. 55).
\bibitem{Bohigas 1983}
 O. Bohigas, R.U. Haq, and A. Pandey, in
 {\em Nuclear Data for Science and Technology}, K.H. B\"ochhoff Ed.
 (Reidel, Dordrecht, 1983), p.809. 
\bibitem{Haq Pandey Bohigas 1982}
 R. U. Haq, A. Pandey, and O. Bohigas, 
 {\em Phys. Rev. Lett.} {\bf 48}, 1086 (1982).
\bibitem{Bohigas Haq Pandey 1985}
 O. Bohigas, R. U. Haq, and A. Pandey, 
 {\em Phys. Rev. Lett.} {\bf 54}, 1645 (1985).
\bibitem{Porter Rosenzweig 1960 Suomalaisen}
 C. E. Porter and N. Rosenzweig, 
 {\em Suomalaisen Tiedeakatemian Toimituksia (Ann. Acad. Sci. Fennicae)}
 {\bf AVI}, No. 44 (1960).
\bibitem{Dyson 1962 140}
 F. J. Dyson, 
 {\em J. Math. Phys.} {\bf 3}, 140 (1962).
\bibitem{Dyson 1962 157}
 F.J. Dyson, 
 {\em J. Math. Phys.} {\bf 3} (1962) 157.
\bibitem{Dyson 1962 166}
 F. J. Dyson, 
 {\em J. Math. Phys.} {\bf 3}, 166 (1962).
\bibitem{Dyson 1970} 
 F. J. Dyson, 
 {\em Commun. Math. Phys.} {\bf 19}, 235 (1970).
\bibitem{Dyson Mehta 1963}
 F. J. Dyson and M. L. Mehta, 
 {\em J. Math. Phys.} {\bf 4}, 701 (1963).
\bibitem{Mehta Dyson 1963}   
 M. L. Mehta and F. Dyson, 
 {\em J. Math. Phys.} {\bf 4}, 713 (1962).
\bibitem{Mehta 1971}
 M. L. Mehta, 
 {\em Commun. Math. Phys.} {\bf 20}, 245 (1971).
\bibitem{Pandey 1983}
 A. Pandey and M.L. Mehta, 
 {\em Commun. Math. Phys.} {\bf 87}, 449 (1983).
\bibitem{Mehta 1984}
 M. L. Mehta and A. Pandey, 
 {\em J. Phys. A: Math. Gen.} {\bf 16}, 2655, L601 (1984).
\bibitem{Harney 1982}
 H. L. Harney, A. Richter, and H. A. Weidenm\"uller, 
 {\em Rev. Mod. Phys.} {\bf 58}, 607 (1986).
\bibitem{Balian 1968}
 R. Balian, 
 {\em Nuov. Cim.} {\bf 57}, 183 (1968).
\bibitem{Gade 1993}
 R. Gade, 
 {\em Nucl. Phys.} {\bf B 398}, 499 (1993).
\bibitem{Andreev 1994}
 A. V. Andreev, B. D. Simons and N. Taniguchi, 
 {\em Nucl. Phys.} {\bf B 432}, 487 (1994).
\bibitem{Haake 1990}
 F. Haake, 
 {\em Quantum Signatures of Chaos} 
 (Springer-Verlag, Berlin, Heidelberg, New York, 1990) 
 Chapters 1, 3, 4, 8, pp 1-11, 33-77, 202-213.
\bibitem{Guhr 1998} 
 T. Guhr T, A. M\"uller-Groeling, and H. A. Weidenm\"uller,
 Phys. Rept. {\bf 299}, 189 (1998). 
\bibitem{Mehta 1990 0}
 M. L. Mehta, 
 {\em Random matrices} 
 (Academic Press, Boston, 1990), Chapters 1, 2, 9, pp 1-54, 182-193.
\bibitem{Reichl 1992}
 L. E. Reichl, 
 {\em The Transition to Chaos In Conservative Classical 
 Systems: Quantum Manifestations} 
 (Springer-Verlag, New York, 1992), 
 Chapter 6, p. 248.
\bibitem{Bohigas 1991}
 O. Bohigas, 
 {\em Proceedings of the Les Houches Summer School on Chaos and
 Quantum Physics}, (North-Holland, Amsterdam, 1991), p.89.
\bibitem{Porter 1965}
 C.E. Porter, 
 {\em Statistical Theories of Spectra: Fluctuations} 
 (Academic Press, New York, 1965).  
\bibitem{Brody 1981}
 T. A. Brody, J. Flores, J. B. French, P. A. Mello, 
 A. Pandey and S. S. M. Wong, 
 Rev. Mod. Phys. {\bf 53}, 385 (1981).
\bibitem{Beenakker 1997}
 C. W. J. Beenakker, 
 Rev. Mod. Phys. {\bf 69}, 731 (1997).  
\bibitem{Feynman 1972}
 R. P. Feynman, 
 {\em Statistical mechanics. A set of lectures.}
 (W. A. Benjamin, Inc., Reading, Massachusetts, 1972).
\bibitem{Tolman 1967}
 R. C. Tolman, 
 {\em The Principles of Statistical Mechanics}
 (Oxford University Press, London, 1967).
\bibitem{Huang 1963}
 K. Huang, 
 {\em Statistical mechanics} 
 (John Wiley and Sons, Inc., New York, 1963).
\bibitem{Anselm 1973}
 A. I. Anselm, 
 {\em Foundations of Statistical Physics and Thermodynamics}
 (Nauka, Moscow, 1973).
\bibitem{Gyarmati 1970}
 I. Gyarmati, 
 {\em Non-equilibrium Thermodynamics; Field Theory and Variational Principles}
 (Springer-Verlag, Berlin, Heidelberg, New York, 1970).
\bibitem{Zubaryev 1971}
 D. N. Zubaryev, 
 {\em Nonequilibrium Statistical Thermodynamics}
 (Nauka, Moscow, 1971), 
 (translation: Consultants Bureau, Inc., New York, 1974).
\bibitem{Fowler 1936}
 R. H. Fowler, 
 {\em Statistical Mechanics}
 (Cambridge University Press, London, 1936).
\bibitem{Fowler 1965}
 R. H. Fowler, E. A. Guggenheim,
{\em Statistical Thermodynamics}
(Cambridge University Press, London, 1965).
\bibitem{Hill 1956}
 T. L. Hill, 
 {\em Statistical Mechanics; Principles and Selected Applications}
 (McGraw-Hill Book Company, Inc., New York, Toronto, London, 1956).  
\bibitem{Landau 1951}
 L. Landau, E. Lifshitz,
 {\em Statistical Physics}
 (Nauka, Moscow, 1951).
\bibitem{Toda 1983}
 M. Toda, R. Kubo, and N. Sait\^{o},
 {\em Statistical Physics, Vol. I Equilibrium Statistical Mechanics;
 Statistical Physics, Vol. II Nonequilibrium Statistical Mechanics}
 (Springer-Verlag, Berlin, Heidelberg, New York, Tokyo, 1983-1985). 
\bibitem{Balescu 1975}
 R. Balescu,
 {\em Equilibrium and Nonequilibrium Statistical Mechanics}
 (John Wiley and Sons, Inc., New York, 1975).
\bibitem{Reiff 1965}
 F. Reiff,
 {\em Statistical and Thermal Physics}
 (McGraw-Hill Book Company, Inc., New York, Toronto, London, 1965).
\bibitem{Balian 1982}
 R. Balian,
 {\em Du microscopique au macroscopique, Cours de physique statistique
 de l'\'ecole polytechnique, Vols. 1, 2}
 (\'Ecole Polytechnique and \'Edition Marketing, Paris, 1982).
\end{thebibliography}
\end{document}